\documentclass[12pt,preprint]{aastex}
\begin{document}
\newcommand{\up}[1]{\ifmmode^{\rm #1}\else$^{\rm #1}$\fi}
\newcommand{\zdot}{\makebox[0pt][l]{.}}
\newcommand{\upd}{\up{d}}
\newcommand{\uph}{\up{h}}
\newcommand{\upm}{\up{m}}
\newcommand{\ups}{\up{s}}
\newcommand{\arcd}{\ifmmode^{\circ}\else$^{\circ}$\fi}
\newcommand{\arcm}{\ifmmode{'}\else$'$\fi}
\newcommand{\arcs}{\ifmmode{''}\else$''$\fi}

\title{The Araucaria Project. Infrared TRGB distances to the Carina and
Fornax dwarf spheroidal galaxies
\footnote{Based on observations obtained with the ESO NTT for 
program 074.D-0505(B)}
}
\author{Grzegorz Pietrzy{\'n}ski}
\affil{Universidad de Concepci{\'o}n, Departamento de Astronomia, 
Casilla 160-C, Concepci{\'o}n, Chile}
\affil{Warsaw University Observatory, Al. Ujazdowskie 4, 00-478, Warsaw, Poland}
\authoremail{pietrzyn@astrouw.edu.pl}
\author{Marek, G{\'o}rski}
\affil{Warsaw University Observatory, Al. Ujazdowskie 4, 00-478, Warsaw, Poland}
\authoremail{mgorski@astrouw.edu.pl}
\author{Wolfgang Gieren}
\affil{Universidad de Concepci{\'o}n, Departamento de Astronomia, 
Casilla 160-C, Concepci{\'o}n, Chile}
\authoremail{wgieren@astro-udec.cl}
\author{Valentin D. Ivanov}
\affil{European Southern Observatory, Ave. Alonso de Cordova 3107, Casilla
19001, Santiago 19, Chile }
\author{Fabio Bresolin}
\affil{Institute for Astronomy, University of Hawaii at Manoa, 2680 Woodlawn 
Drive,  Honolulu HI 96822, USA}
\authoremail{bresolin@ifa.hawaii.edu}
\author{Rolf-Peter Kudritzki}
\affil{Institute for Astronomy, University of Hawaii at Manoa, 2680 Woodlawn 
Drive, Honolulu HI 96822, USA}
\authoremail{kud@ifa.hawaii.edu}
\begin{abstract}
We present distance determinations for two Local Group dwarf spheroidal galaxies,
Carina and Fornax, based on the near-infrared magnitudes of the tip of the 
red giant branch (TRGB). For Carina we derive true distance moduli of 20.09 and
20.13 mag in the J and K bands, respectively, while
for Fornax the same distance modulus of 20.84 mag was derived in
both filters. The statistical errors
of these determinations are of order  0.03-0.04 mag, whereas the systematic
uncertainties on the distances are 0.12 mag in the J band and 0.14 mag
in the K band. The distances obtained from the near-infrared TRGB method 
in this paper agree very well with those obtained for these two galaxies
from optical calibrations of the TRGB method, 
their horizontal branches, RR Lyrae variables, 
and the near-infrared magnitudes of their red clumps.
\end{abstract}


\section{Introduction}
The main goal of the Araucaria project is to improve the calibration
of the cosmic distance scale from accurate observations of the various
primary stellar distance indicators in nearby galaxies (e.g. Gieren et
al. 2005b). In the course of our project we are observing Cepheids, 
RR Lyrae stars, red clump stars, blue supergiants, eclipsing binaries
and the tip of the red giant branch (TRGB) brightness in both optical 
and infrared (IR) domains.
Since with IR photometry one can minimize 
the influence of interstellar reddening on the derived distances, and 
in many cases also 
the population dependence of the standard candles, this part of our
project is particularly important for precise distance determinations 
to our target galaxies, and therefore for a more accurate calibration of the 
extragalactic distance scale. 
In our previous papers we already demonstrated that 
red clump stars (Pietrzynski and Gieren 2002;
Pietrzynski, Gieren and Udalski 2003), Cepheids (e.g. Pietrzynski et al.
2006; Gieren et al. 2005a, 2006, 2008a, 2008b; Soszynski et al. 2006), 
and RR Lyrae stars (Pietrzynski et al. 2008, Szewczyk et al. 2008)
in the IR domain are very accurate tools for distance determination 
to nearby galaxies. 
In this paper we extend our near-infrared distance work 
to the TRGB method in the J and K bands, starting with
the two dwarf spheroidals Carina and Fornax.
The success of the TRGB method as a tool for distance measurement begun  in
1993 when   Lee et al. (1993) convincingly showed
that  the optical I band magnitude of the TRGB does not depend on metallicity
and age in environments with metallicity lower than about - 0.7 dex.
Subsequently, very detailed studies of a possible
population dependence of the I band TRGB magnitude  (e.g. Kennicutt et al. 1998
; Ferrarese et al. 2000; Udalski, 2000)
confirmed  the  finding of Lee et al. and converted
the I-band TRGB method into a widely used standard candle.
Due to its simplicity and the relatively large absolute brightness 
of stars at the tip
of the red giant branch in the I band (about -4 mag),
over the past decade this method has been
applied to most of the nearby galaxies (e.g. Ferrarese et
al., 2000, Karatchentsev et al., 2003).
 However, a very important problem in
this technique was the potentially strong influence of reddening on the
distances obtained from the optical data, especially for galaxies located at
low Galactic latitudes.

Very recently important progress has been achieved  by Ivanov and Borissova (2002) and 
 Valenti, Ferraro and Origlia (2004), who have provided an accurate calibration 
of the J, H and K band absolute magnitudes of the TRGB.
Because of the fact that the TRGB magnitude in the IR domain is much brighter
than in the optical (${\rm M}_{\rm k}$ $<$ -6 mag), is very insensitive
to reddening (which in the near-IR domain is an order
of magnitude smaller than in the optical), and that the infrared calibration
is valid over a very broad range of metallicities ( -2.2 $<$ [Fe/H] $<$ -0.4 dex)
these works opened  the  possibility to use the infrared TRGB magnitude
to measure accurate distances to galaxies located out to several Mpc.
The technique has been already applied to derive distances to 
several nearby galaxies (e.g. Cioni et al. 2000 (Magellanic Clouds), 
Rejkuba 2004 (NGC 5128)).  

Our paper is  organized as follows. In the next section we describe the  
observations, and the reduction and calibration methods we used. Then 
we present the distance determination to the Fornax and Carina Local Group galaxies,
followed by a discussion of the errors associated to our results
and a comparison with the distances previously derived for these galaxies
from other techniques.  Finally we present a summary and final remarks.

\section{Observations, Data Reduction and Calibration}
The near infrared data presented in this paper were collected as a
part of the Araucaria Project, with the ESO NTT
telescope on La Silla equipped with the SofI infrared camera
(Moorwood, A., Cuby, J.G., Lindman, C., 1998). 
The Large Field Mode, with a field of view of 4.9 x 4.9 arcmin, and a
scale of 0.288 arcsec/pixel was used.
During one night 9 and 4 SofI fields were observed in Carina
and Fornax, respectively (see Table 1 and Figs. 1 and 2). 
These galaxies were already subject of earlier deep IR imaging performed
by our group with the ESO VLT
(Pietrzynski, Gieren and Udalski 2003). However, those observations were 
optimized for studying the relatively faint red clump stars, and the stars 
having brightnesses similar to the TRGB magnitude were saturated in
these images, or fall into the nonlinear regime of the ISAAC camera.
Therefore, in order to measure accurately the K and J band magnitudes of the TRGB 
in Carina and Fornax we decided to observe them again with the SofI 
camera, using very short integrations of 1.5 s. Such short 
exposures guarantee that  high quality photometry can be obtained 
for stars having magnitudes as bright as the expected magnitudes 
of the TRGB in our target galaxies. In order to account for rapid sky level
variations in the infrared
domain, our observations were performed with a dithering technique.
In the K and J band  filters, we  averaged over 20 (Carina) and  10
(Fornax) consecutive 1.5 s
integrations (DITs)  at any given
pointing before moving the telescope to a randomly selected different
position within 20$\times$20 arcsec square. 12 such dithering positions 
were sufficient to  extract the sky and 
obtain obtain photometry deep enough to measure accurately the TRGB
magnitude in both filters. 

The reductions were performed in a similar manner to those
described in Pietrzy{\'n}ski and Gieren (2002). The sky was subtracted
from the images with a two-step process implying masking
of the stars with the $xdimsum$ IRAF package. Then, the individual
images for each field and filter were flatfielded and stacked into a
final composite image.
The PSF photometry was carried out with the DAOPHOT and ALLSTAR
programs.
About 10-20 relatively bright and isolated stars were selected visually,
and the first PSF model was derived from them. Then,
following Pietrzy{\'n}ski, Gieren and Udalski (2002),  we iteratively
improved
the PSF model by subtracting all stars from their neighbourhood and
re-calculating the
PSF model. After three such iterations no further improvement was noted,
and the corresponding PSF model was  adopted as the final one. 

Since the observations were performed under non-photometric conditions 
we decided to transform our data onto the standard system using the 
2MASS data. Typically about 30 stars from the 2MASS Point Source Catalog
(Wachter et al. 2003) 
were found in a given SofI field. The scatter (rms) of the calculated 
zero point offsets was always smaller than 0.01 mag.
Several of the observed fields overlap with the regions observed by 
Pietrzy{\'n}ski, Gieren and Udalski (2003). Therefore it was possible to
check the zero point of our photometry transformed onto the 2MASS system
with the carefully calibrated deep photometry obtained by these authors.
In each case the 
difference in the corresponding zero points was found to be smaller 
than 0.02 mag.
We therefore conclude that the zero point uncertainty of our photometry 
does not exceed 0.02 mag.

\section{Distance determination}
From the K, J-K color-magnitude diagrams for Carina and Fornax obtained
from our data (Fig. 3), the  stars on the RGB were selected and their 
respective luminosity functions were calculated using a
bin size of 0.08 mag, which represents a reasonable compromise between
the number of stars in a magnitude bin, and the magnitude resolution we 
can achieve. 
In each galaxy the location of the TRGB is well marked (see Figure 4)
and does not depend neither on bin size nor on the starting point
of the histogram.

In order to obtain TRGB magnitudes in a more objective way we 
computed the Gaussian-smoothed luminosity functions and use the Sobel 
edge-detection filter, following the procedure described in detail 
by Sakai, Madore and Freedman (1996). The resulting luminosity functions
and the outputs of the edge-detection filter for the J and K bands 
and both galaxies  are presented in Figs. 5 and 6.
As can be seen, the location of the highest peak in the edge-detection filter output,
which we interpreted as the TRGB magnitude, is very well defined 
in both filters, in each galaxy. The automatically detected 
TRGB locations agree very well with those obtained from the visual
inspection of the binned luminosity functions.
The presence of  AGB stars adds additional noise in the procedure 
of the TRGB analysis, and if the intermediate-age population is very large  in 
a given galaxy the TRGB detection may in principle  be  difficult. 
This effect was studied in detail by Makarov et al. (2006), 
and Barker et al. (2004), who conlcuded that the TRGB detection 
is quite insensitive to the AGB contamination.
As can be seen in Figs. 5 and 6, 
the locations of the most significant peaks in the Sobel 
filter outputs, and therefore the corresponding J or K band 
magnitudes of the TRGB, could be unambiguously identified
in both galaxies.

We would like to notice that our K band TRGB magnitude of 
14.45 $\pm$ 0.04 mag differs by some 3 sigma from the 
results  obtained by Gullieuszik et al. 
(2007; 14.59 $\pm$ 0.03). This difference is most probably caused 
by the different technique of TRGB detection 
employed by these authors (i.e., the Maximum Likelihood 
Algorithm (MLA) of Makarov et al. 2006). Rizzi et al. (2006),
applying both Sobel filter and MLA techniques to the same HST data set 
of NGC 300, obtained results which in some cases were different by 0.08 mag.

In Table 2 we report the J and K band magnitudes of
the TRGB in the Carina and Fornax galaxies obtained with the 
Sobel filter technique.
The Galactic foreground reddenings toward the galaxies 
were estimated from the Schlegel, Finkbeiner and Davis (1998) extinction
maps and are E(B-V)=0.03 for Fornax, and E(B-V)=0.06 for Carina.
Following Pietrzy{\'n}ski, Gieren and Udalski (2003), we adopted for the
(mean) metallicities of the RGB the values -1.0 dex for Fornax (Saviane,
Held and Bertelli 2000; Tolstoy et al. 2002), and -1.7 dex for Carina 
(Koch et al. 2006, Koch et al. 2008). These adopted reddening and metallicity
values are also given in Table 2.

In order to derive the infrared TRGB distances, the empirical calibration of
Valenti, Ferraro and Origlia (2004) which was obtained from extensive near-infrared
observations of 24 Galactic globular clusters covering the wide
metallicity range from -2.12 to -0.49 dex was used (equations 1 and 2). 
It was demonstrated by these authors that these calibrations agree very
well with the predictions from theoretical models.

${\rm {M}_{J}^{TRGB} = -5.67 - 0.31 \times [Fe/H]}$ (1)\\

${\rm {M}_{K}^{TRGB} = -6.98 - 0.58 \times [Fe/H]}$ (2)\\

Adopting  the metallicities and reddenings listed in Table 2, and our
measured TRGB magnitudes, the following true distance moduli were calculated from
these calibration equations:

Carina: 20.09 $\pm$ 0.03  (J band),  20.13 $\pm$ 0.04 (K band)\\

Fornax: 20.84 $\pm$ 0.03 (J band), 20.84 $\pm$ 0.04 (K band) \\

For both galaxies, the respective distances from the J and K bands agree within the
statistical uncertainties of the TRGB magnitudes.

\section{Discussion}
With the assumed uncertainty of the photometric J and K band zero points 
of 0.02 mag, an estimated error associated to the extinction determinations
of 0.02 mag, and an assumed uncertainty in the adopted metallicities
of 0.2 dex we calculate the total systematic uncertainties of our
distance moduli determinations for Carina and Formax to be of 0.12 mag 
and 0.14 for the J and
K band filters, respectively, for both galaxies. The dominant part in
the systematic uncertainties of the current infrared TRGB distances
to Carina and Fornax comes from the assumed uncertainties on the appropriate
metallicities of the red giant branch tip stars. Their effect on the
distance has been estimated from the metallicity coefficients in equations
1 and 2. It should be noted here that we did not take into account
 any contribution from possible systematic errors in the
coefficients themselves in the calibration of  Valenti, Ferraro, and Origlia 
(2004). As our final distance results, we adopt a true distance modulus of
(20.11 $\pm$ 0.13) mag for the Carina dSph galaxy, and (20.84 $\pm$ 0.15) mag
for the Fornax dSph galaxy.

Results of previous distance determinations to the Fornax and Carina galaxies
from different methods reported in the literature are given
in Table 3.
Pietrzy{\'n}ski, Gieren and Udalski (2003) obtained from deep K band  imaging
of red clump stars true distance moduli of  20.165 $\pm$ 0.015 mag and 20.858
$\pm$ 0.013 mag for Carina and Fornax, respectively. Similar results
were obtained for the Carina distance from an analysis of its RR Lyrae 
stars ( 20.10 $\pm$ 0.12 mag; Dall'Ora et al. (2003)), from its optical TRGB and
horizontal branch (HB) 
luminosities (20.05 $\pm$ 0.06 mag and 20.12 $\pm$ 0.08 mag; Smecker-Hane
et al. 1994), and from dwarf Cepheids (20.06 $\pm$ 0.12 mag; Mateo et al. 1998).  
A shorter distance to Carina of 19.94 mag was calculated by Udalski 
(2000) from I band photometry of red clump stars and the TRGB, 
and from V band photometry of RR Lyrae stars. However, Udalski's distance
estimation for Carina
was tied to a short assumed LMC distance of 18.24 mag, which is probably
an underestimation of the true LMC distance modulus (e.g. Schaefer 2008,
Fouqu{\'e} et al. 2007). Correcting the LMC distance to a more likely value
near 18.5 would bring Udalski's result for the Carina distance in close
agreement with the present work, and the other results cited above.

Regarding the Fornax dSph galaxy, optical photometry  of the TRGB and HB 
has yielded the 
following distance moduli: 20.76 $\pm$ 0.02 (Buonanno et al. (1999), 
20.70 $\pm$ 0.12 (Saviane et al. (2000), 20.65 $\pm$ 0.11 (Bersier
(2000), and 20.71 $\pm$ 0.07 (Rizzi et al. 2007). Greco et al. (2005) 
 measured $(m-M)_{0}$ = 20.72 $\pm$ 0.10 for this galaxy from RR Lyrae stars.
A slightly shorter distance modulus (20.66 mag) was obtained by Bersier 
from I band photometry of  red clump stars.
Recently Gullieuszik et al. (2007) derived the distance modulus for Fornax
from near-infrared photometry of the TRGB (20.75 $\pm$ 0.19) and red clump
stars (20.74 $\pm$ 0.11) applying population corrections derived from
models. Our current distance result for Fornax is at the upper limit
of the distribution of these earlier measurements, but within its uncertainty
clearly consistent with these. It is likely that the quality of the
distance determinations to galaxies from their TRGB in the optical I band
and in the near-IR JK bands is quite similar as to the correction
of the effect of the varying metallicities of different red giant
populations; however, a decisive advantage of the infrared TRGB method 
is its basic insensitivity to reddening corrections. While this has no
strong consequences in the case of the Carina and Fornax dwarf galaxies
due to their low foreground reddenings, and (assumed) negligible internal
reddenings, things might be very different when highly reddened galaxies
are under study.

\section{Summary and Conclusions}
>From near-infared photometry of the Carina and Fornax dSph galaxies
we have derived
the distances to these galaxies using the calibration of the TRGB
absolute magnitudes in the J and K bands given by Valenti, Ferraro and Olivia
(2004). We have obtained the following results:

\noindent 20.09 $\pm$ 0.03 $\pm$ 0.12 mag (Carina, J band) \\
20.14 $\pm$ 0.04 $\pm$ 0.14 mag (Carina, K band) \\
20.84 $\pm$ 0.03 $\pm$ 0.12 mag (Fornax, J band) \\
20.84 $\pm$ 0.04 $\pm$ 0.14 mag (Fornax, K band) \\

Our distance determinations are in very good agreement with the 
results obtained from deep infrared imaging  of red clump stars, and an
analysis of the optical photometry of the TRGB and HB,  and RR Lyrae stars
in these galaxies. 
Our results show that the K and J band magnitudes of the TRGB in sufficiently
metal-poor galaxies are promising tools for distance determinations. 
However, in order to further check and potentially improve this technique 
it should be applied to measure distances to other
galaxies showing a broad range of environments.
 A comparison of such results with the distances measured from other distance
indicators from both optical and near infrared photometric data
should shed more light on the population dependence of 
the stellar distance indicators, and also on the problems related to the
interstellar extinction corrections.  Curently our group is involved in 
near infrared imaging of several other nearby galaxies which will enable
such comparative studies.

\acknowledgments
WG and GP gratefully acknowledge financial support for this
work from the Chilean Center for Astrophysics FONDAP 15010003, and from
the BASAL Centro de Astrofisica y Tecnologias Afines (CATA). 
Support from the Polish grant N203 002 31/046 and the FOCUS 
subsidy of the Fundation for Polish Science (FNP)
is also acknowledged. It is a great pleasure
to thank the support astronomers at ESO-La Silla
for their expert help in the observations. 
Finally, we would like to thank the anonymous referee
for  very constructive suggestions, which helped us
to substantially improve our paper.

\begin{figure}[p]
\vspace*{20cm}
\includegraphics{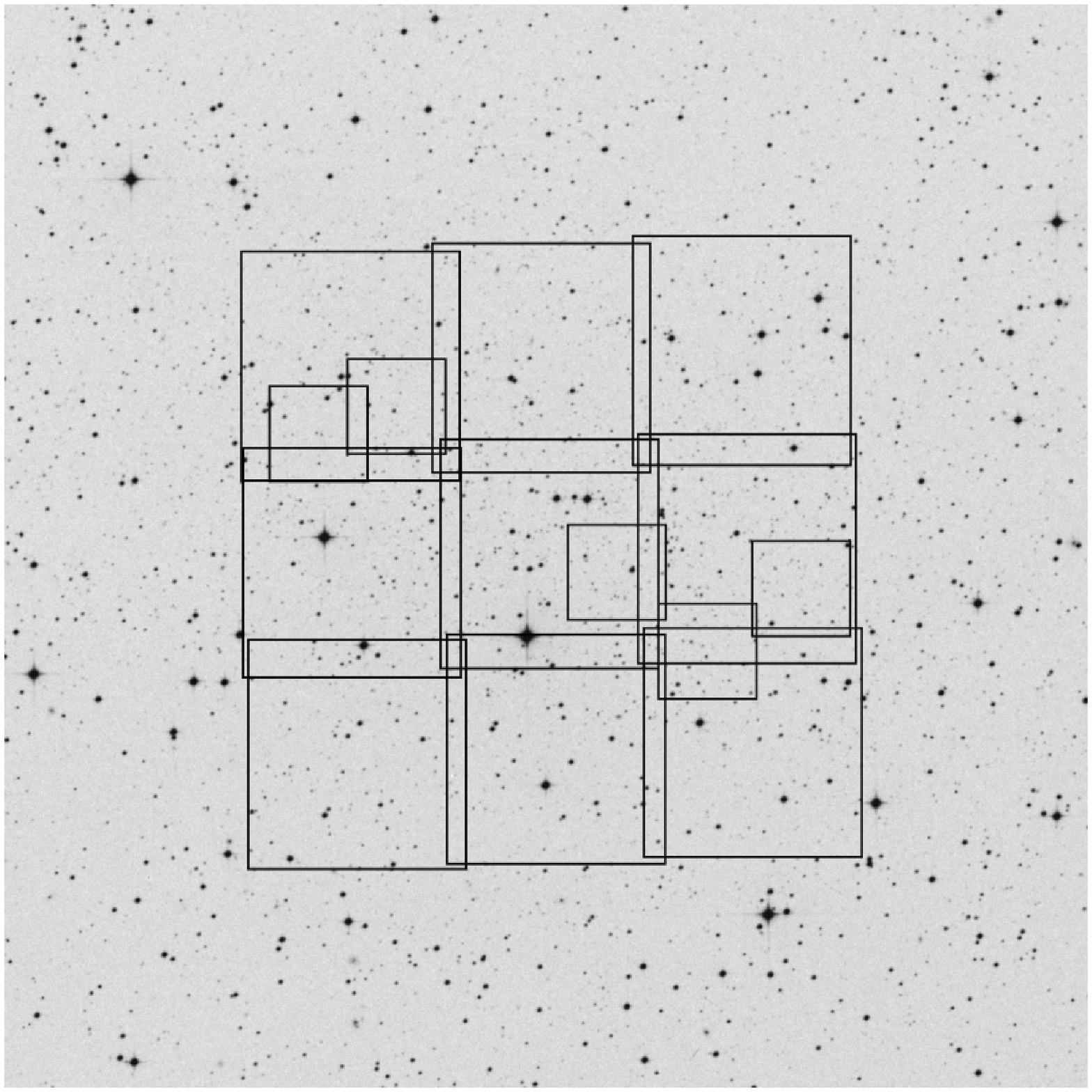}
\caption{The nine  5 x 5 arcmin NTT/SofI fields observed in the Carina dSph galaxy,
marked on the 20 x 20 arcmin DSS image of this galaxy (large squares). 
The location of the five 2.5 x 2.5 arcmin VLT/ISAAC fields observed previously 
by Pietrzynski et al. (2003) are  shown with small squares.
North is up and East to the left.
}
\end{figure}

\begin{figure}[p]
\vspace*{20cm}
\includegraphics{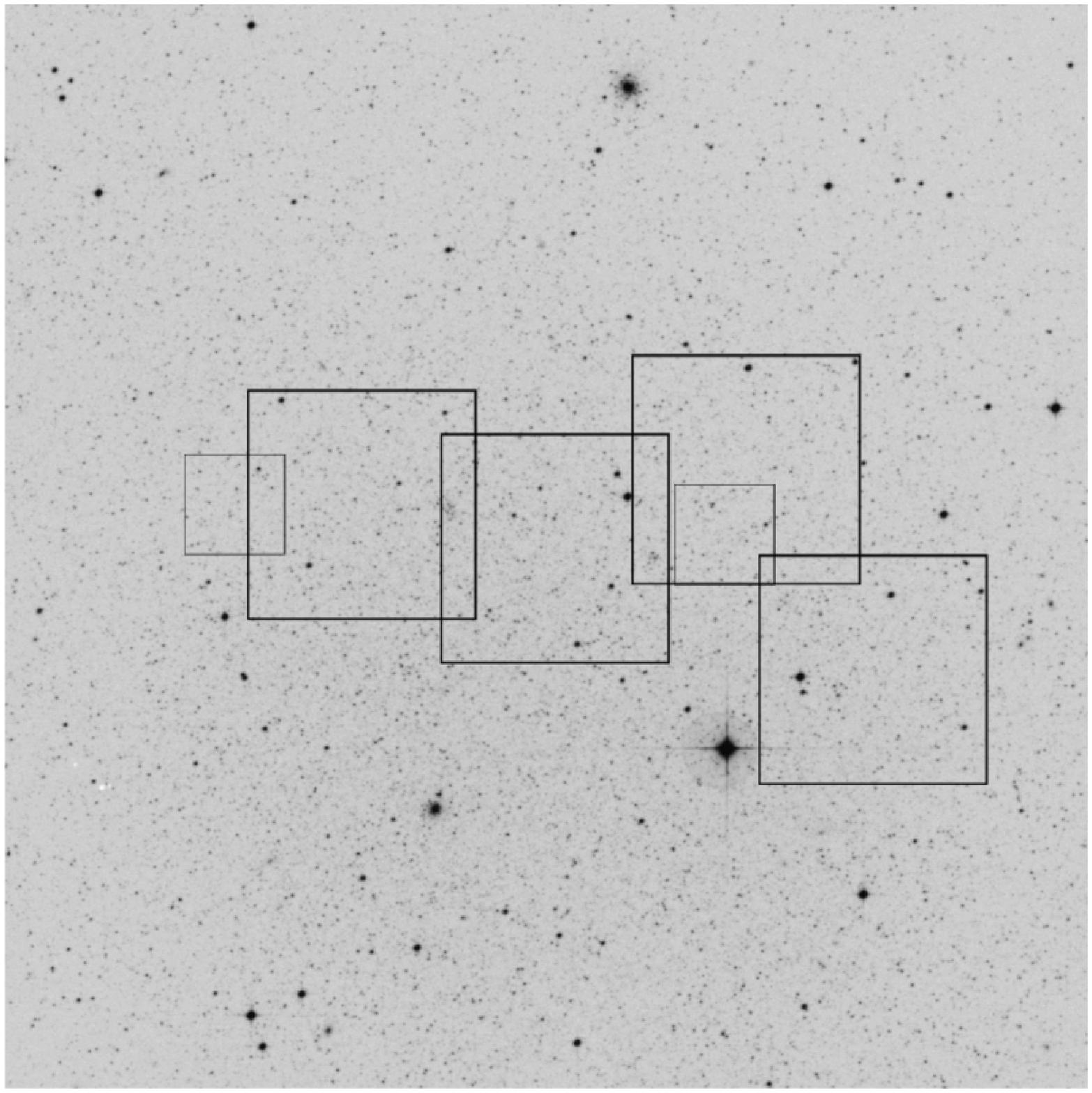}
\caption{
The four  5 x 5 arcmin NTT/SofI fields observed in the Fornax dSph galaxy, 
marked on the 20 x 20 arcmin DSS image of this galaxy (large squares). 
The location of the two 2.5 x 2.5 arcmin VLT/ISAAC fields observed previously
by Pietrzynski et al. (2003) are shown  with small squares.
North is up and East to the left.
}
\end{figure}

\begin{figure}[p]
\vspace*{18cm}
\includegraphics{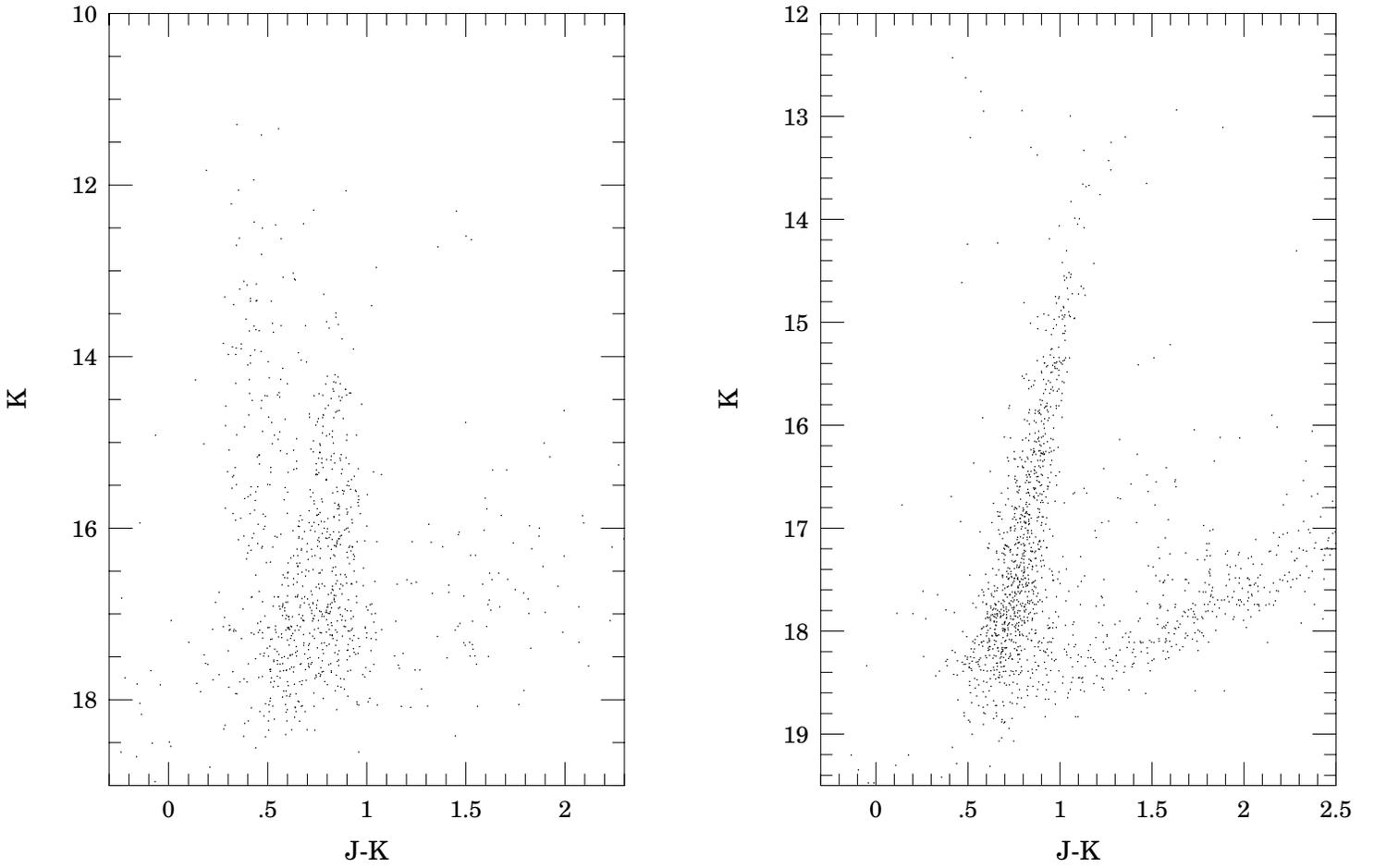}
\includegraphics{fig3b.ps}
\caption{K, J-K color-magnitude diagrams for the Carina (left panel) and Fornax 
(right panel) dwarf spheroidal galaxies obtained
from our data.}
\end{figure}

\begin{figure}[p] 
\vspace*{18cm}
\includegraphics{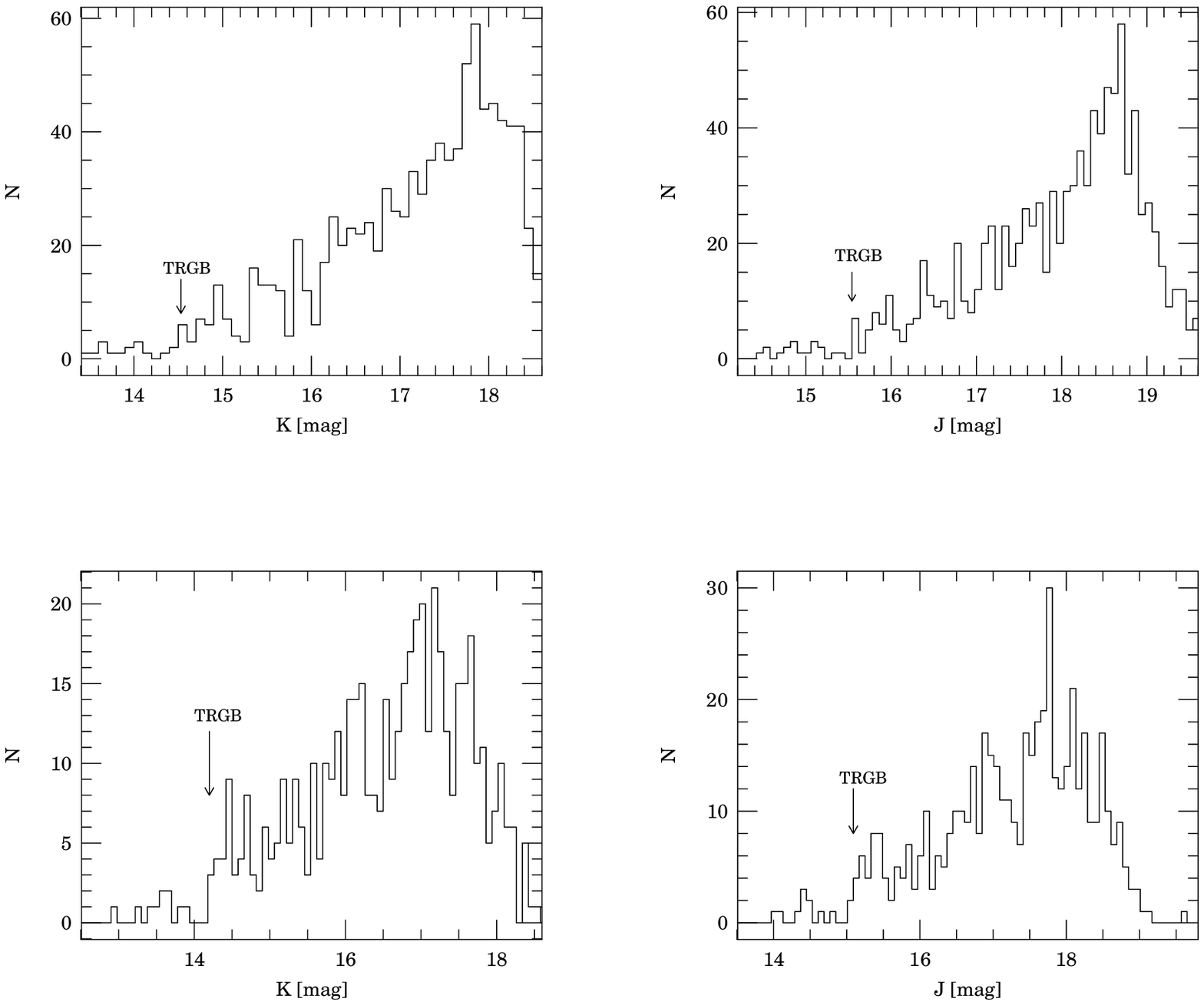}
\caption{The J and K band luminosity functions of the red giant branch 
in the Carina dSph (two lower panels) and Fornax galaxies (two upper
panels), constructed from our SOFI data. The arrows points at the
position of the respective TRGB magnitudes. }
\end{figure}  

\begin{figure}[p]
\vspace*{18cm}
\includegraphics{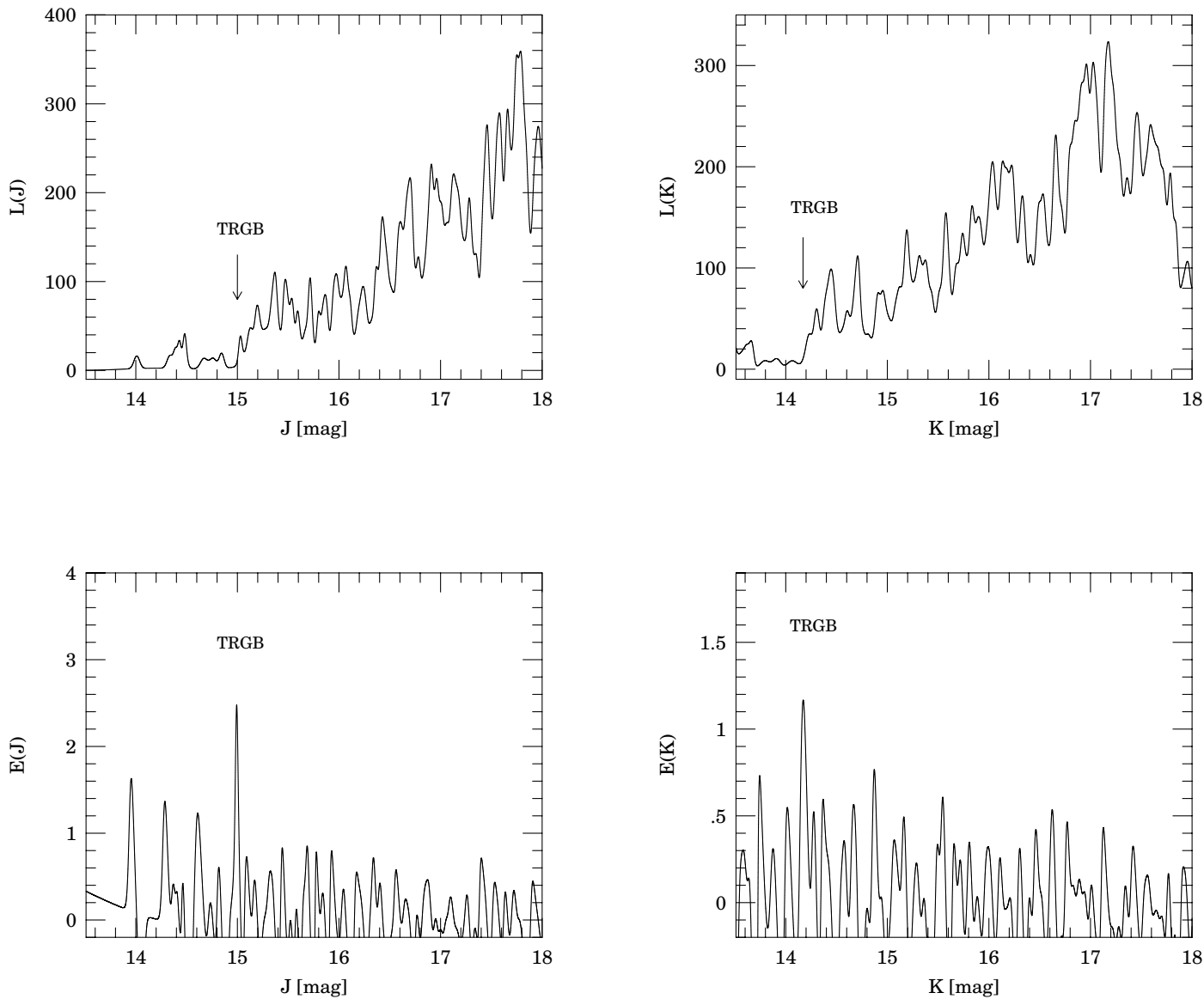}
\caption{The J and K band Gaussian-smoothed  luminosity functions of the red giant branch
in the Carina dSph (top) and the corresponding outputs of the
edge-detection filter (bottom). 
}
\end{figure}

\begin{figure}[p]
\vspace*{18cm}
\includegraphics{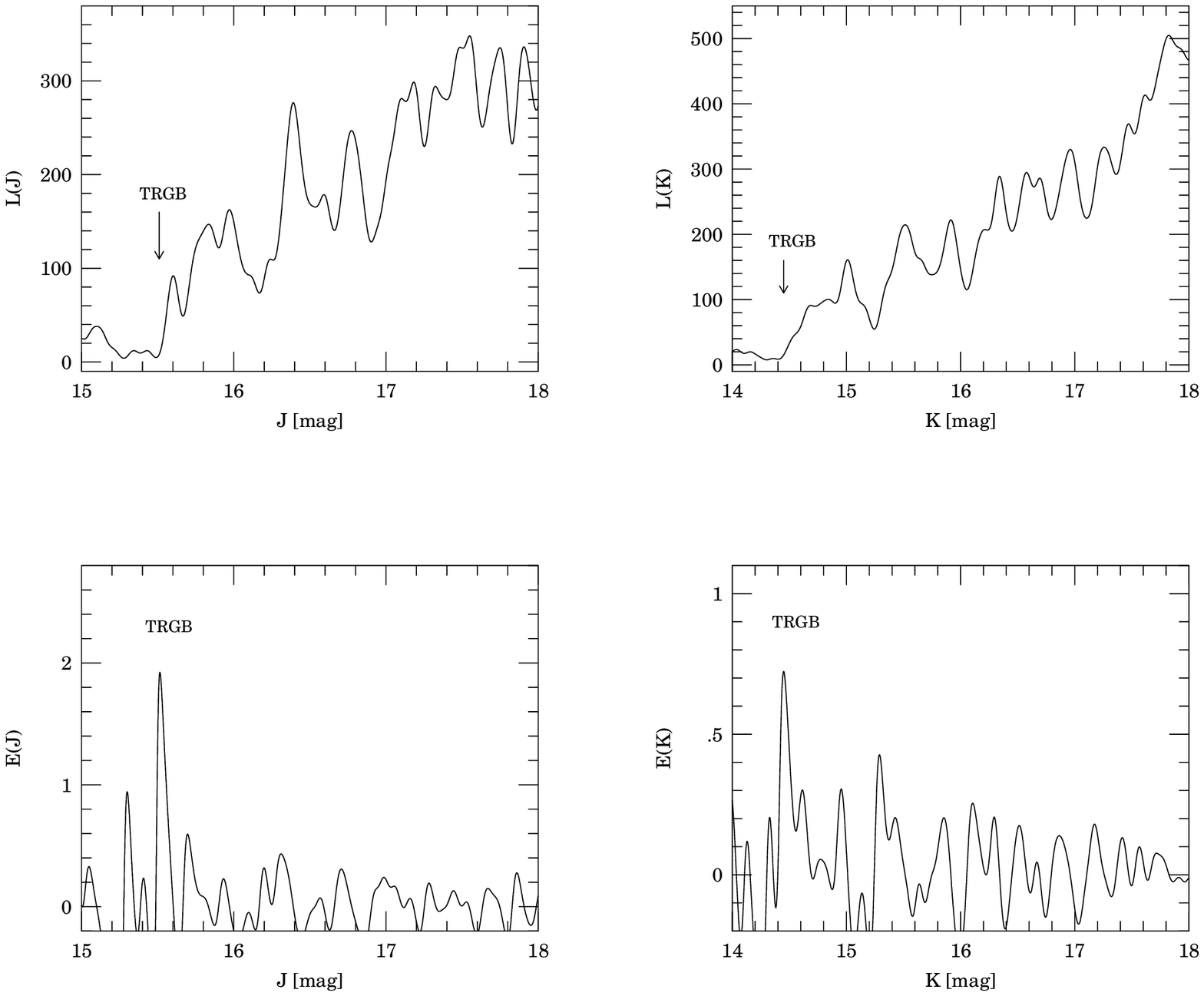}
\caption{The same as in Figure 5, but for the Fornax dwarf galaxy.
}
\end{figure}

\clearpage
\begin{deluxetable}{ccc}
\tablecaption{Coordinates of the NTT/SofI fields observed in the
Carina and Fornax dwarf spheroidal galaxies}
\tablehead{\colhead{Field name} & \colhead{RA 2000} & \colhead{DEC 2000}}
\startdata
CarI    &      $   06^h 42^m 05^s.4  $  &  $  -50^{\circ} 53' 24''.6 $\\
CarII   &      $   06^h 42^m 05^s.4  $  &  $  -50^{\circ} 57' 58''.2 $\\
CarIII  &      $   06^h 42^m 05^s.4  $  &  $  -51^{\circ} 02' 24''.2 $\\
CarIV   &      $   06^h 41^m 36^s.9  $  &  $  -50^{\circ} 53' 24''.6 $\\
CarV    &      $   06^h 41^m 36^s.9  $  &  $  -51^{\circ} 02' 24''.2 $\\
CarVI   &      $   06^h 41^m 08^s.8  $  &  $  -50^{\circ} 53' 16'' $ \\
CarVII  &      $   06^h 41^m 08^s.2  $  &  $  -50^{\circ} 57' 58''.2 $\\
CarVIII &      $   06^h 41^m 08^s.2  $  &  $  -51^{\circ} 02' 24''.2 $\\
CarIX   &      $   06^h 41^m 36^s.9  $  &  $  -50^{\circ} 57' 58''.2 $\\
ForI    &      $    02^h 40^m 18^s.1  $ & $   -34^{\circ} 25' 02''.3 $\\
ForII   &      $    02^h 39^m 56^s.2  $ & $   -34^{\circ} 26' 04''.4 $\\
ForIII  &      $    02^h 39^m 34^s.5  $ & $   -34^{\circ} 24' 04''.5 $\\
ForIV   &      $    02^h 39^m 20^s.7  $ & $   -34^{\circ} 28' 26''.1 $\\
\enddata
\end{deluxetable}

\begin{deluxetable}{ccccc}
\tablecaption{Observed J and K band magnitudes of the TRGB in the Carina and Fornax
dSph galaxies, together with information about metallicity and reddening}
\tablehead{\colhead{Galaxy} & \colhead{TRGB (K)} & \colhead{TRGB (J)}& \colhead{[Fe/H]} & \colhead{E(B-V)} 
 \\
\colhead{} & \colhead{[mag]} & \colhead{[mag]} &
\colhead{[dex]} & \colhead{[mag]} \\
} 
\startdata
Carina  & 14.17    & 15.00    & -1.7   & 0.06 \\
Fornax  & 14.45    & 15.51    & -1.0   & 0.03 \\ \hline
\enddata
\end{deluxetable}

\begin{deluxetable}{cccccc}
\tablecaption{Distance determinations for the Carina and Fornax dSph galaxies 
obtained with different stellar indicators, in the optical and near-infared 
domain}
\tablehead{\colhead{Galaxy} & \colhead{method} & \colhead{band}&
\colhead{distance modulus} & error & reference \\
 \\
\colhead{} & \colhead{} & \colhead{} &
\colhead{[mag]} & \colhead{[mag]} & \\
}
\startdata
Carina  & red clump    & K    & 20.165   & 0.015 & Pietrzynski et al. (2003)\\
Carina  & RR Lyrae     & V    & 20.10    & 0.12  & Dall'Ora et al. (2003) \\
Carina  & TRGB         & I    & 20.05    & 0.06  & Smecker-Hane et al. (1994) \\
Carina  & HB           & I    & 20.12    & 0.08  & Smecker-Hane et al. (1994)\\
Carina  & DC           & V    & 20.06    & 0.12  & Mateo et al. (1998)\\
Fornax  & TRGB         & I    & 20.76    & 0.20  & Buonanno et al. (1999) \\
Fornax  & HB           & I    & 20.70    & 0.12  & Rizzi et al. (2007) \\
Fornax  & RR Lyrae     & V    & 20.72    & 0.10  & Greco et al. (2005) \\
Fornax  & red clump    & I    & 20.66    &  --   & Bersier (2000) \\
Fornax  & TRGB         & I    & 20.65    & 0.11  & Bersier (2000) \\
Fornax  & TRGB         & K    & 20.75    & 0.19  & Gullieuszik et al. (2007) \\
Fornax  & red clump    & K    & 20.858   & 0.013 & Pietrzynski et al. (2003)\\
\hline
\enddata
\end{deluxetable}


\begin{references}
\reference{} Barker, M.K., Sarajedini, A., Harris, J., 2004, \apj, 606, 869

\reference{} Bersier, D., 2000, \apjl, 534, 23

\reference{} Buonanno, R., Corsi, C.E., Castellani, M., Marconi, G.,
Fusi Pecci, F., Zinn, R., 1999, \aj, 118, 1671

\reference{} Cioni, M.R., van der Marel, R.P., Loup, C. and Habing, H.J., 2000, \aap, 359, 601

\reference{} Dall'Ora, M., Ripepi, V., Caputo, F., et al., 2003, \aj, 126, 197

\reference{} Ferrarese, L., et al., 2000, \apjs, 128, 431

\reference{} Fouqu{\'e}, P., Arriagada, P., Storm, J., Barnes, T.G., Nardetto, N., Merand, A.,
Kervella, P., Gieren, W., Bersier, D., Benedict, G.F. and McArthur, B.E., 2007,
\aap, 476, 73

\reference{} Gieren, W., Pietrzy{\'n}ski, G., Soszy{\'n}ski, I., Bresolin, F., 
Kudritzki, R.-P., Minniti, D., and Storm, J., 2005a, \apj, 628, 695

\reference{} Gieren, W., Pietrzy{\'n}ski, G., Bresolin, F., et al., 2005b, 
Messenger, 121, 23

\reference{} Gieren, W., Pietrzy{\'n}ski, G., Nalewajko, K., Soszy{\'n}ski, I., 
Bresolin, F., Kudritzki, R.P., Minniti, D., and Romanowsky, A., 2006, \apj, 647, 1056

\reference{} Gieren, W., Pietrzy{\'n}ski, G., Soszy{\'n}ski, I., Bresolin, F.,
Kudritzki, R.P., Storm, J. and Minniti, D., 2008a, \apj, 672, 266

\reference{} Gieren, W., Pietrzy{\'n}ski, G., Szewczyk, O., Soszy{\'n}ski, I.,
Bresolin, F., Kudritzki, R.P., Urbaneja, M.A., Storm, J. and Minniti, D., 2008b,
\apj, 683, 611

\reference{} Greco, C., Clementini, G., Held, E.V., et al., 
 2005, Resolved Stellar Populations, Cancun Mexico. 
 
\reference{} Gullieuszik, M., Held, E.V., Rizzi, L., Saviane, I., Momani, Y.,
and Ortolani, S., 2007, \aap, 467, 1025

\reference{} Ivanov, V.D., Borissova, J., 2002, \aap, 390, 937

\reference{} Karatchentsev, I., et al., 2003, \aap, 404, 93

\reference{} Kennicutt, R.C.Jr., et al., 1998, \apj, 498, 181

\reference{} Koch, A., Grebel, E.K., Wyse, R.F.G., et al., 2006, \aj, 131, 895

\reference{} Koch, A., Grebel, E.K., Gilmore, G.F., et  al., 2008, \aj, 135, 1580

\reference{} Lee, M.G., Freedman, W.L., Madore, B.F., 1993, \apj, 417, 553

\reference{} Makarov, D., Makarova, L., Rizzi, L., et al., 2006, \aj, 132, 2729

\reference{} Mateo, M., Olszewski, E., W., Vogt, S.S., Keane, M.J.,
1998, \aj, 116, 2315

\reference{} Moorwood, A., Cuby, J.G., Lidman, C., 1998, ESO Messenger,
94, 7

\reference{} Pietrzy{\'n}ski, G., and Gieren, W., 2002, \aj, 124, 2633

\reference{} Pietrzy{\'n}ski, G., Gieren, W., and Udalski, A., 2002, \pasp, 114, 298

\reference{} Pietrzy{\'n}ski, G., Gieren, W., and Udalski, A., 2003, \aj, 125, 2494

\reference{} Pietrzy{\'n}ski, G., Gieren, W., Soszy{\'n}ski, I., Bresolin, F., 
Kudritzki, R.-P.,Dall'Ora, M., Storm, J., and Bono, G., 2006, \apj, 642, 216

\reference{} Pietrzynski, G., Gieren, W., Szewczyk, O., Walker, A.,
Rizzi, L., Bresolin, F., Kudritzki, R.-P., Nalewajko, K., Storm, J.,Dall'Ora, M., 
Ivanov, V., 2008, \aj, 135, 1993

\reference{} Rejkuba, M., 2004, A\&A, 413, 903

\reference{} Rizzi, L., Bresolin, F., Kudritzki, R.P., Gieren, W., 
Pietrzynski, G., 2006, \apj, 638, 766

\reference{} Rizzi, L., Held, E.V., Saviane, I., Tully, R.B.,
Gullieuszik, M., 2007, \mnras, 380, 1255

\reference{} Sakai, S., Madore, B., Freedman, W.L., 1996, \apj, 461, 713

\reference{} Saviane, I., Held, E.V., Bertelli, G., 2000, \aap, 355, 56

\reference{} Schaefer, B.E., 2008, \aj, 135, 112

\reference{} Schlegel, D.J., Finkbeiner, D.P., and Davis, M., 1998,
\apj, 500, 525

\reference{} Smecker-Hane, T.A., Stetson, P.B., 
Hesser, J.E., Lehnert, M.D., 1994, \aj, 108, 507


\reference{} Soszy{\'n}ski, I., Gieren, W., Pietrzy{\'n}ski, G.,
Bresolin, F.,
Kudritzki,R.P., and Storm, J., 2006, \apj, 648, 375

\reference{} Szewczyk, O., Pietrzynski, G., Gieren, W., Storm, J.,  Walker, A.,
Rizzi, L., Kinemuchi, K.,  Bresolin, F., Kudritzki, R.-P., 
Dall'Ora, M., 2008, \aj, 136, 272
 
\reference{} Tolstoy, E., et al., 2002, \apss, 281, 217 

\reference{} Udalski, A., 2000, Acta Astron., 50, 279

\reference{} Valenti, E., Ferraro, F.R., Origlia, L., 2004, \mnras, 354, 815

\reference{} Wachter, S., Hoard, D.W., Hansen, K.H., Wilcox, R.E.,
Taylor, H.M., Finkelstein, S.L., 2003, \apj, 586, 1356
\end{references}
\end{document}